\documentclass{aa520}
\usepackage{graphicx,natbib,amsmath,amssymb,fnpara}
\usepackage{txfonts}
\bibpunct[, ]{(}{)}{,}{a}{}{,}

\begin{document}

   \title{The X-ray afterglows of gamma-ray bursts GRB\,001025A and
          GRB\,010220 observed with \emph{XMM-Newton}}
   \titlerunning{X-ray afterglows of GRBs observed with \emph{XMM-Newton}}

   \author{D.~Watson\inst{1}
          \and
          J.~N.~Reeves\inst{1}
          \and
          J.~Osborne\inst{1}
          \and
          P.~T.~O'Brien\inst{1}
          \and
          K.~A.~Pounds\inst{1}
          \and
          J.~A.~Tedds\inst{1}
          \and
          M.~Santos-Lle\'{o}\inst{2}
          \and
          M.~Ehle\inst{2}
          }

   \offprints{D.~Watson, email: \texttt{wat@star.le.ac.uk}}

   \institute{X-ray Astronomy Group, Dept. of Physics and Astronomy, University of Leicester, Leicester LE1 7RH, UK
         \and
             XMM-Newton Operations Centre, European Space Agency, Vilspa, Apartado 50727, 28080 Madrid, Spain
             }

   \date{Received ; accepted }

   \abstract{The X-ray afterglows of GRB\,001025A and GRB\,010220 were detected
             by \emph{XMM-Newton} with an average 0.2--10.0\,keV flux of 4.4
             and $3.3\times10^{-14}$\,erg\,cm$^{-2}$\,s$^{-1}$ respectively;
             the afterglow of GRB\,001025A is observed to decay.  Afterglows
             at other wavelengths were not detected for either burst.  A set
             of broadened soft X-ray emission lines are detected in the
             afterglow of GRB\,001025A, at $5.0\,\sigma$ significance above
             a Galactic-absorbed power-law continuum.  The spectra of both
             afterglows are significantly better fit by a variable abundance
             thermal plasma model than by an absorbed power-law and are
             consistent with the observations of GRB\,011211, indicating
             that thermal emission from light elements may be common in the
             early X-ray afterglows of GRBs.
   }

   \maketitle

\section{Introduction\label{introduction}}
Much of the recent progress in the understanding of gamma-ray bursts (GRBs)
has come from bursts detected with good spatial accuracy with
\emph{BeppoSAX} and it is particularly at X-ray wavelengths that GRB
afterglows are detected \citep{2001grba.conf...97P}, about half producing no detectable optical
afterglow emission \citep{2001A&A...369..373F}.  It is also only at X-ray wavelengths that
emission lines are detected in afterglows, allowing firm estimates to be
made of the cosmological redshifts and the outflow velocities of the
afterglow material \citep{2000Sci...290..955P,2002Natur.416..512R}.
\emph{XMM-Newton} \citep{2001A&A...365L...1J}, with its
large effective area, is particularly suited to this work.  Previous
detections of emission lines in GRB afterglows with \emph{BeppoSAX} and
\emph{Chandra} have concentrated on emission from highly-ionised iron
\citep{1998A&A...331L..41P,2000Sci...290..955P,2000ApJ...545L..39A,2001ApJ...557L..27Y};
however recent observations with \emph{XMM-Newton} have revealed several
emission lines at lower energies \citep{2002Natur.416..512R}.

Most plausible mechanisms for the production of a GRB involve a newly-formed
black hole surrounded by a short-lived accretion disk regardless of the progenitor
\citep{1998ApJ...494L..45P,%
1999ApJ...524..262M,1999A&A...344..573R,2001Sci...291...79M}. Recent
evidence suggests that the progenitors of long-duration GRBs are massive
stars \citep{1998ApJ...494L..45P,1999Natur.401..453B,%
1999ApJ...524..262M,2002Natur.416..512R}.

Various models have been proposed to account for the emission spectra (in
particular the claim of a high equivalent width Fe emission line
\citep{2000ApJ...545L..39A}) and lightcurves of the afterglow.  For instance,
the nearby reprocessor model
\citep{2000ApJ...545L..73R,2001ApJ...559L..83B,astro-ph/0110654}
involving reflection of synchrotron emission from the walls of a cone
tunnelled out of a massive star, yielding large equivalent width Fe emission
lines; or a `supranova' model \citep{1998ApJ...507L..45V} which invokes a
time delay ($\gtrsim30$ days) between an initial supernova (SN) explosion and the
GRB, giving a spectrum briefly dominated either by the recombination of Fe
in a photoionised plasma \citep{1998ApJ...507L..45V} or reflection of
synchrotron emission off the walls of a wide funnel excavated in the
SN remnant \citep{2001ApJ...550L..43V}.  Recently,
\citet{2002Natur.416..512R} have suggested that the early X-ray afterglow
spectrum of GRB\,011211 is dominated by thermal emission from a
metal-enriched, but notably Fe-poor collisionally-ionised plasma ejected in
a recent SN explosion and heated by the GRB.

In Sect.~\ref{observations} we report on observations of two GRB afterglows
with \emph{XMM-Newton}, presenting the spectra in Sect.~\ref{results}.  In
Sect.~\ref{discussion} these results are discussed and their implications
for other X-ray observations of afterglows examined. Our conclusions are in
Sect.~\ref{conclusions}.  Unless otherwise stated, all errors quoted are 90\%
confidence limits for one parameter of interest.  A cosmology where
H$_0=75$\,km\,s$^{-1}$\,Mpc$^{-1}$ and q$_0=0.5$ is assumed throughout.

\section{Observations and data reduction\label{observations}}

\subsection{GRB\,001025A}
\citet{2000GCN...861....1S} and \citet{2000GCN...863....1H} report detection
of a GRB by the RXTE all-sky monitor, NEAR and \emph{Ulysses} beginning at
03:10:05~UT on 25 October 2000.  It had a duration of $\sim5$\,s, a fluence
of $3.2\times10^{-6}$\,erg\,cm$^{-2}$ and a peak flux of
$2.0\times10^{-6}$\,erg\,cm$^{-2}$\,s$^{-1}$ in the 25--100\,keV band as
detected by \emph{Ulysses}.  No variable source optical counterpart was
detected down to an R-band magnitude limit of $\sim24.5$
\citep{2000GCN...867....1F}.

\emph{XMM-Newton} began observing the error-box of GRB\,001025A 45 hours
after the burst for two EPIC-pn exposures of 8\,ks and 16\,ks and single
37\,ks exposures of the MOS cameras in full frame mode.  The thin
filters were used in each case.

The data were processed and reduced with the SAS, version 5.2, datasets from
both EPIC-MOS cameras were co-added and both EPIC-pn exposures were also
co-added and the resulting two datasets fit simultaneously.  Source
extraction regions were 35\arcsec\ in radius and an off-source background
extraction region of 70\arcsec\ radius was chosen.  Both single and double
pattern (as well as triple and quadruple pattern for the MOS), good (FLAG=0)
events were used with the ready-made response
matrices\footnote{\tt http://xmm.vilspa.esa.es/ccf/epic} provided by the
\emph{XMM-Newton} SOC.  The spectra were binned with a minimum of 20 counts
per bin.  The final X-ray source positions were determined after
cross-correlation with the USNO A2.0 optical catalogue based on the SAS task
\emph{eposcorr} \citep[see][]{Tedds:2000}.

Two sources were detected in the IPN error-box
\citep{2000GCN...884....1A,2000GCN...869....1A}; the brighter, with
coordinates (J2000) R.A. 08h36m35.86s,  Dec. $-13$\degr 04\arcmin
12.28\arcsec,\ and a 68\% error radius of 0.6\arcsec,\ was seen to decay slowly (at
99.8\% confidence), a power-law decay ($F\propto t^{-\beta}$) with index
$\beta=3.0\pm1.9$ fitting the lightcurve well.
We identify this source as the afterglow of GRB\,001025A.  The Galactic hydrogen
absorbing column in this direction is $6.1\times10^{20}$\,cm$^{-2}$
\citep[using the FTOOL \emph{nh}]{1990ARA&A..28..215D}.  Its mean 0.2--10.0\,keV
flux was $4.4\times10^{-14}$\,erg\,cm$^{-2}$\,s$^{-1}$, corresponding to an
afterglow luminosity of
$2.8\times10^{43}$\,erg\,s$^{-1}$ using the redshift ($z=0.53$) determined
from the thermal plasma model fit to the data (see
Sect.~\ref{001025A_results}). The second source is fainter
($1.2\times10^{-14}$\,erg\,cm$^{-2}$\,s$^{-1}$) and shows no evidence for
variability.

\subsection{GRB\,010220}
GRB\,010220 was detected  by \emph{BeppoSAX} at 22:51:07~UT on 20 February
2001 and subsequently localised to within a 4\arcmin\ radius circle; the burst
duration was $\sim40$\,s, with $\sim660$\,count\,s$^{-1}$ peak flux in the
40--700\,keV band \citep{2001GCN...956....1M}.  No optical counterpart was
detected to a limiting magnitude of R$\simeq23.5$
\citep{2001GCN...958....1B}, however the burst position is in the Galactic
plane ($b^{II} = 1.41$), where the Galactic nebula IC\,1805 lies along the line
of sight \citep{2001GCN...957....1C}.  The Galactic hydrogen absorbing
column is $8.6\times10^{21}$\,cm$^{-2}$ \citep[using the FTOOL \emph{nh}]{1990ARA&A..28..215D}.  At 14.8 hours after the burst,
\emph{XMM-Newton} began observing at the coordinates of the \emph{BeppoSAX}
error-circle.  The observation of GRB\,010220 was contaminated by a high and
variable background.  To mitigate this effect, only data where the
background rate was relatively low were used.  The EPIC-pn exposure was
43\,ks; screening for flares left 20\,ks of good data.  Full frame mode and
the medium filter were used for the observation.  No data were available from
the EPIC-MOS cameras.

The data were processed and reduced with the SAS version 5.2.  A source
extraction region of 20\arcsec\ radius and an off-source background
extraction region of 80\arcsec\ radius were chosen.  Both single and double
pattern, good (FLAG=0) events were used with the ready-made response
matrix$^1$ provided by the \emph{XMM-Newton} SOC.  The spectra were binned
with a minimum of twenty counts per bin.  Source positions were determined
as noted above.

Four sources were detected in the \emph{BeppoSAX} error circle; the
brightest source is 18\arcsec\ from the centre, with the others at
$>3$\arcmin.  No variability is detected in any of the sources at greater
than $2\sigma$ confidence.  We assume that the brightest source, at
coordinates (J2000): R.A.  02h37m01.66s, Dec.\
+61\degr45\arcmin56.01\arcsec\ with a 68\% confidence error circle of
1.2\arcsec\ radius is the GRB afterglow.  This source decays at 94\%
confidence, with a best-fit decay index, $\beta=1.2\pm1.0$.  Its mean
0.2--10.0\,keV flux was $3.3\times10^{-14}$\,erg\,cm$^{-2}$\,s$^{-1}$
(corresponding to an afterglow luminosity of
$7.3\times10^{43}$\,erg\,s$^{-1}$ in the rest frame 0.2--10.0\,keV band
based on the best-fit redshift ($z=1.0$) from the thermal plasma fit to the
data).

\section{Spectral fitting\label{results}}
In order to test the wider applicability of the collisionally-ionised plasma
model proposed by \citet{2002Natur.416..512R} to explain the
\emph{XMM-Newton} observations of the afterglow of GRB\,011211, our new GRB
X-ray afterglow data were fit with the same set of models.  They were {\bf
a)} a power-law, {\bf b)} a power-law with a variable, cold, redshifted
absorber, {\bf c)} a variable abundance collisionally-ionised plasma model
\citep[the MEKAL model,][]{1985A&AS...62..197M,1995ApJ...438L.115L} with the abundances of Mg,
Si, S, Ar and Ca fit jointly, Ni allowed to vary freely and all other
elements fixed at the solar value and {\bf d)} an ionised reflection model
\citep{2001ApJ...559L..83B} where the emission arises purely from the X-ray
flux scattered off material with twice the solar elemental abundance, also
modified by a cold, redshifted absorber. In all cases an absorber fixed at
the Galactic value was included.  

In order to test the significance of the fit improvements, 10\,000 spectra
were simulated using the parameters derived from fitting an absorbed
power-law (model {\bf b}) to the data in each case.  Models {\bf b)} and
{\bf c)} were fit to the simulated data and the difference in
$\chi^2$ computed.  The results of these tests are reported in
Table~\ref{fit_comparisons}.  Using a random gaussian distribution of
initial parameters centred on the parameters derived from fitting an
absorbed power-law to the data, a second set of 10\,000 simulated spectra
was generated.  This second set of simulations yielded results consistent
with the first.

\begin{table}
 \caption{$\chi^2$/degrees of freedom for model fits to both GRB afterglows.
          Model {\bf a)} is a power-law.  Model {\bf b)}, is the same as
          model {\bf a)} modified by a cold, redshifted absorber. {\bf c)} is a
          variable abundance collisionally ionised plasma model with the
          abundances of Mg, Si, S, Ar and Ca fit jointly, Ni allowed to vary
          freely and all other elements fixed at the solar value.  Model
          {\bf d)} is an ionised reflection model where the emission arises
          purely from the X-ray flux scattered off material with twice the
          solar elemental abundance, also modified by a cold, redshifted
          absorber.  In all cases an absorber fixed at the Galactic value
          was included.  Column four is the percentage of
          spectral fits (simulated using model {\bf b}) with
          greater $\chi^2$ differences between models {\bf b)} and {\bf c)} 
          than the real data.}
  \label{fit_comparisons}
 \begin{tabular}{@{}lccccc@{}}
\hline\hline
  GRB	& a		& b		&	MC Prob.	& c	& d\\
\hline
001025A	& 95.7/66 	& 68.6/64	& 	0.13\%		& 54.1/63	& 65.5/63\\
010220	& 15.4/15	& 14.8/14	& 	0.16\%		& 5.2/13	& 6.7/12\\
\hline
 \end{tabular}
\end{table}

\begin{table}
 \caption{Best-fit parameters from spectral fits to the afterglow data.  In
          the upper section, parameters of the fit to the collisionally
          ionised plasma model (model {\bf c} in Sect.~\ref{results}) are
          presented.  Parameters of the best fit to an absorbed power-law
          (model {\bf b}) are included for reference in the lower section. 
          Values in parentheses are 90\% confidence limits for one
          interesting parameter.}
 \label{mekal_fits}
 \begin{tabular}{@{}lcccc@{}}
\hline\hline
GRB	& T (keV)	& $z$	& Mg/Si/S/Ar/Ca & Ni \\
	&		&	& \multicolumn{2}{c}{(Solar Abundances)}\\
\hline
001025A	& 3.4	& 0.53	& 2.0	& 26\\
	& (2.9--3.9)	& (0.50--0.55)	& (0.7--3.7)	& (14--40)\\
010220	& 6.0	& 1.0	& 10	& 95\\
	& (3.6--12.7)	& (0.97--1.07)	& frozen	& (32--248)\\[6pt]
\hline
\hline
	& $\Gamma$	& $z$	& N$_{\rm H}$ ($10^{21}$\,cm$^{-2}$)\\
\hline
001025A & 2.50   & 0.5  & 3\\
	& (2.26--2.97)	& (0--7.1)  & (1--184)\\
010220  & 2.1   & 1.0   & 16\\
	& (1.5--3.1)   & frozen  & (0--67)\\
\hline
 \end{tabular}
\end{table}

\subsection{GRB\,001025A}
\label{001025A_results}

The spectrum for this afterglow is significantly better fit by the thermal
plasma model than the absorbed power-law
(Table~\ref{fit_comparisons})---only 13 in 10\,000 simulations are as good a
fit.  While the continuum shape dictates that the plasma temperatures in the
simulated spectra will be consistent with each other and with the fit to the
real data, only four of the redshifts determined from the thirteen best-fit
simulated spectra are within 90\% confidence limits of the value from the
real data, reinforcing the fact that the emission lines are not systematic
deviations.

\begin{figure}
 \includegraphics[angle=-90,width=\columnwidth,clip=]{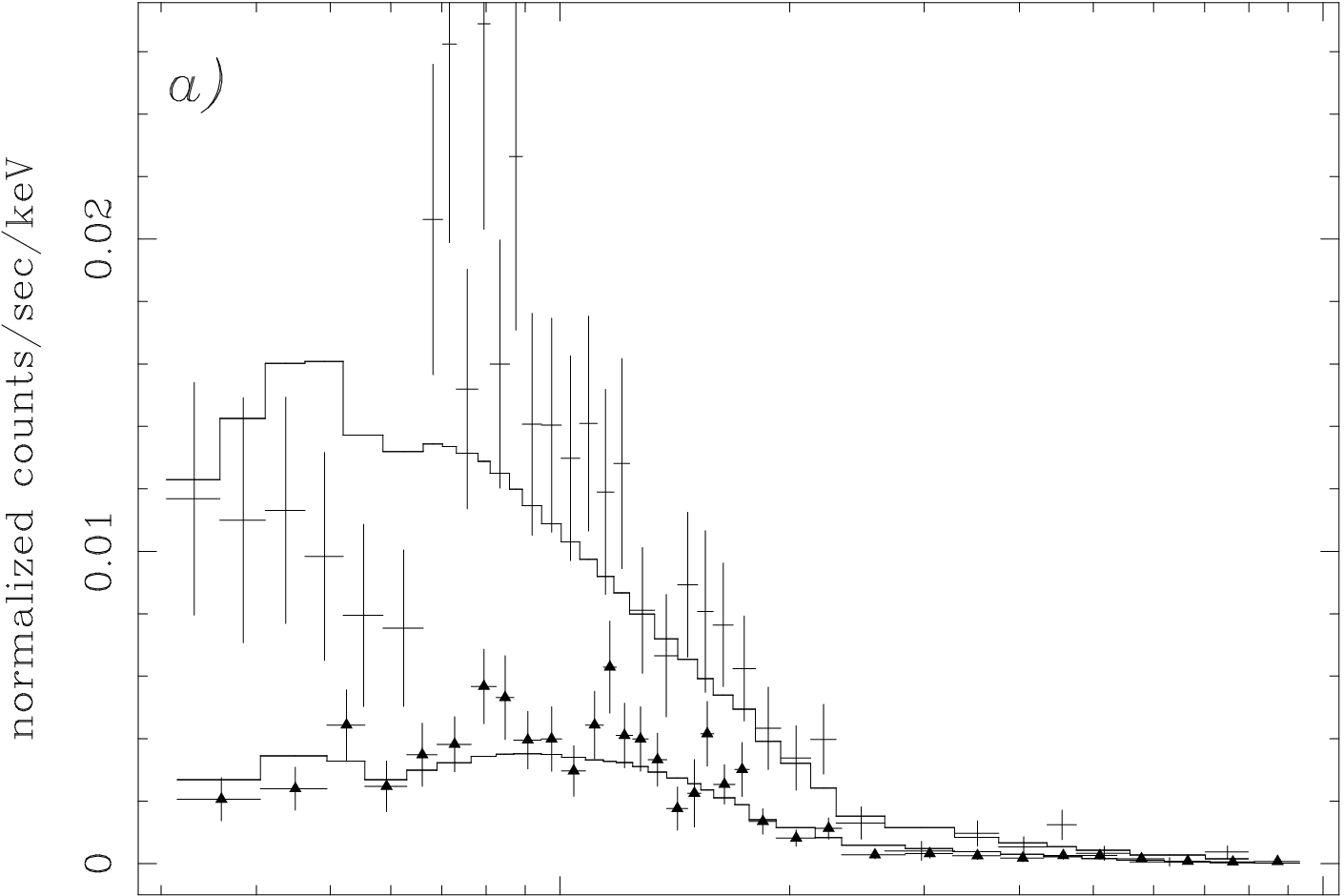}
 \includegraphics[angle=-90,width=\columnwidth,clip=]{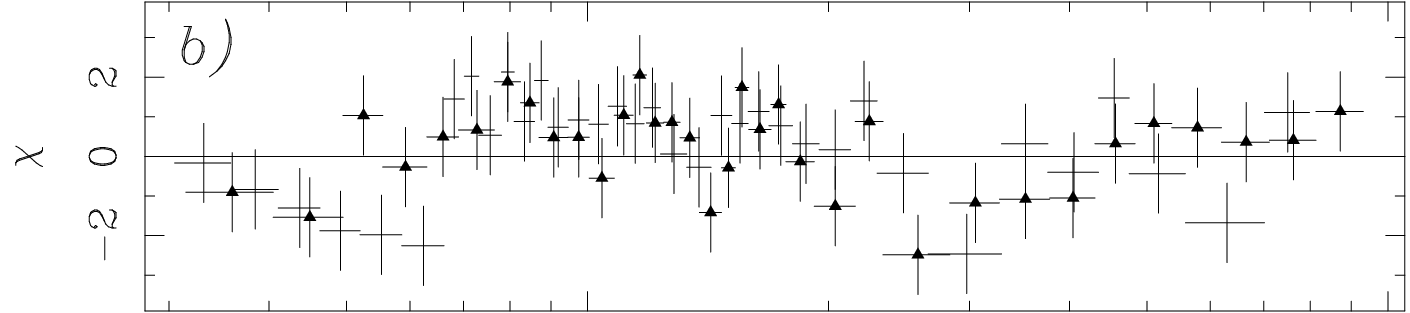}
 \includegraphics[angle=-90,width=\columnwidth,clip=]{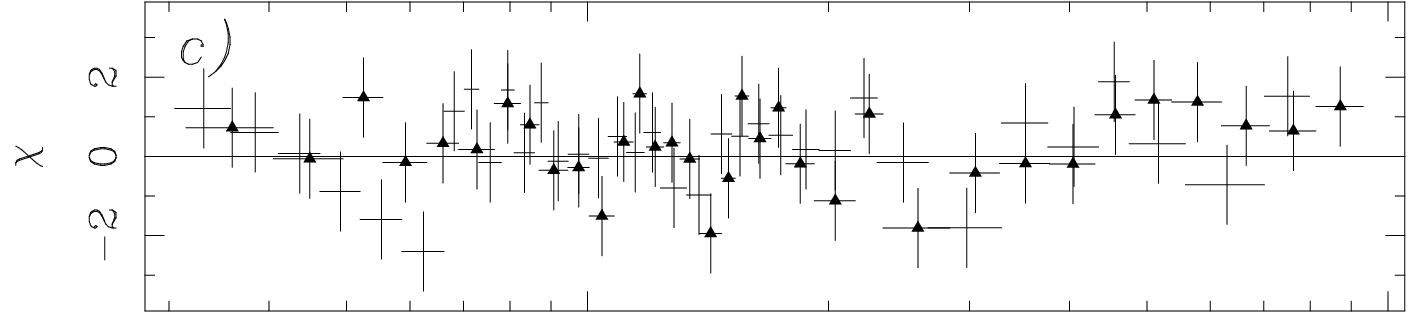}
 \includegraphics[angle=-90,width=\columnwidth,clip=]{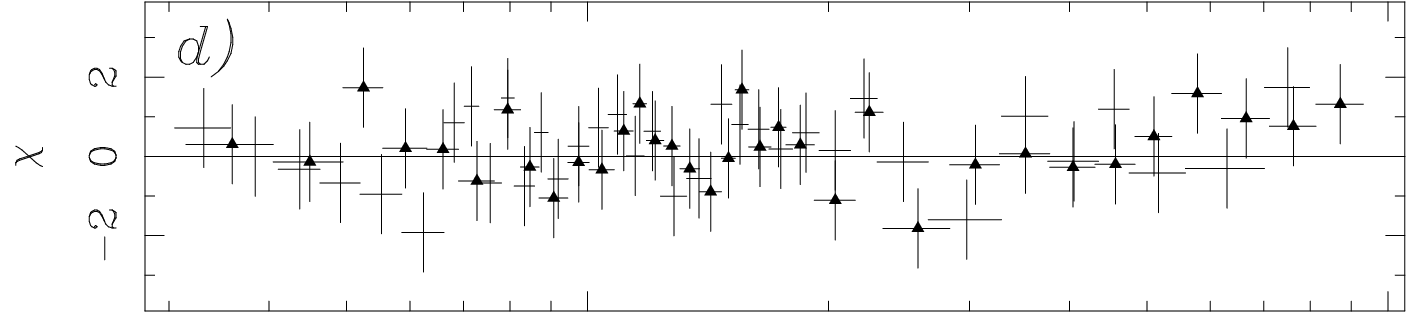}
 \includegraphics[angle=-90,width=\columnwidth,clip=]{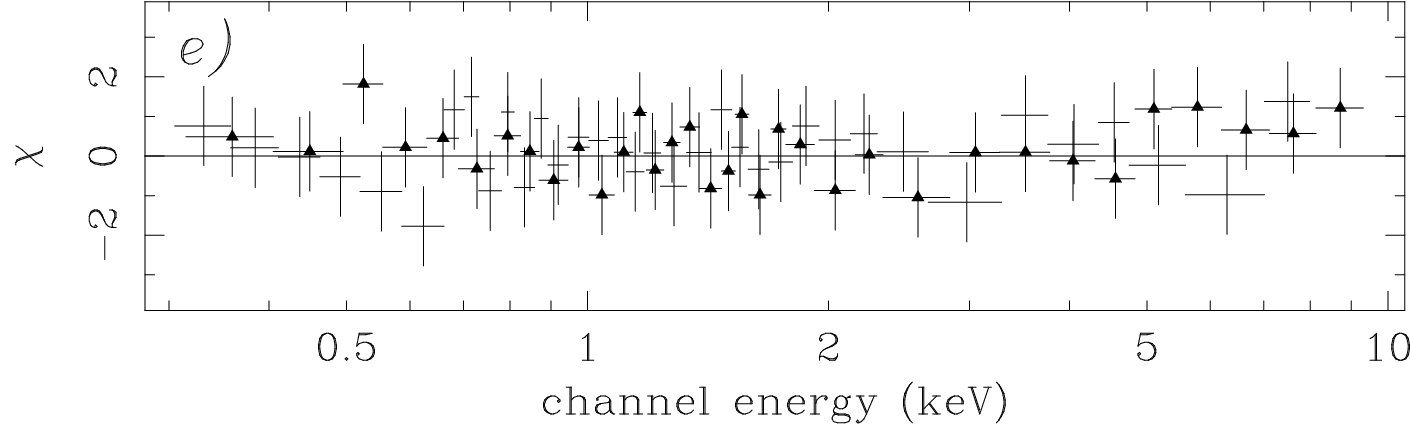}
 \caption{EPIC-pn (crosses) and combined MOS (filled triangles) spectrum of the afterglow of GRB\,001025A;
          \textbf{\textit{a)}} data fit to a Galactic-absorbed power-law model;
          \textbf{\textit{b)}} residuals to plot \textit{a)};
          \textbf{\textit{c)}} fit residuals to a Galactic-absorbed power-law model with variable redshifted absorption;
          \textbf{\textit{d)}} fit residuals to a collisionally-ionised plasma model with variable abundance of Mg, Si,
          S, Ar and Ca and freely variable Ni abundance (VMEKAL), absorbed by the Galactic column;
          \textbf{\textit{e)}} fit residuals to a Galactic-absorbed power-law and five Gaussian emission lines model, allowing the line widths to vary jointly.}
 \label{complete_pn+mos_001025A_spectrum}
\end{figure}

The need for the thermal emission component arises
from the soft excess observed between $\sim0.5$ and 2\,keV
(Fig.~\ref{complete_pn+mos_001025A_spectrum}), which we suggest may be due
to the blend of lines from Mg, Si, S, Ar and Ni-L.  The thermal plasma model
fit to the data yields a temperature of 3.4\,keV and a redshift of 0.53 and
does not require super-solar abundance of light metals
(Table~\ref{mekal_fits}).

In order to parameterise the line emission, a power-law with Galactic
absorption and Gaussian emission lines was fit to the data, allowing the
line widths to vary together.  For five emission lines, the $\chi^2$ is 40.1
for 55 degrees of freedom, (giving a null hypothesis probability of
$3\times10^{-4}$ ($3.6\,\sigma$) over the absorbed power-law and
$5\times10^{-7}$ ($5.0\,\sigma$) over a power-law with Galactic absorption.
Adding lines to a bremsstrahlung continuum yielded similar significances,
($3\times10^{-4}$ ($3.6\,\sigma$) over the absorbed bremsstrahlung and
$2\times10^{-4}$ ($3.8\,\sigma$) over a bremsstrahlung with Galactic
absorption), though the fit was worse, $\chi^2=45.4$ for 55 degrees of
freedom.

The best-fit energies for the lines are: $4.7_{-0.4}^{+0.8}$, $2.2\pm0.1$,
$1.64\pm0.07$, $1.16\pm0.05$ and $0.80_{-0.05}^{+0.04}$\,keV with a FWHM of
$240^{+70}_{-60}$\,eV with individual line significances (null hypothesis
probabilities) of 0.12, 0.02, $8\times10^{-4}$, $8\times10^{-4}$ and
$4\times10^{-4}$ respectively.  Single line significances were determined by
using the best-fit Galactic-absorbed power-law with five Gaussian emission
lines and removing individual lines to assess their f-statistic. 
Probabilities were then determined using the f-test.  The observed equivalent
width is $\sim800$\,eV for the 4.7\,keV line and $\sim400\,$eV for the
others. Fitting the K$\alpha$ emission lines of
\ion{Mg}{xii} (1.46\,keV), \ion{Si}{xiv} (1.99\,keV),
\ion{S}{xvi} (2.60\,keV), \ion{Ar}{xviii} (3.30\,keV) and \ion{$^{56}$Ni}{xxviii} (8.10\,keV), gives a best-fit redshift
of $0.7^{+0.3}_{-0.1}$.  This redshift is within two standard deviations of
the redshift determined from the thermal plasma fit
(Table~\ref{mekal_fits}). Allowing for an outflow velocity of $0.0-0.1\,c$
\citep{2002Natur.416..512R}, we conclude that the redshift of the host
galaxy is likely to lie in the range 0.5--1.2.

It has been suggested that reflection of synchrotron emission from ionised
material could produce the emission lines observed in GRB afterglows
\citep[e.g.][]{astro-ph/0206116}.  An ionised reflection model is as
good a fit to the data as an absorbed power-law;
however, adding a power-law continuum to the reflection component
disimproves the fit significantly, indicating that the afterglow must be
viewed so that little of the continuum is observed, if this fit is to be
acceptable.  This implies a line-of-sight outside the GRB cone, a possibility
ruled out by the detection of the GRB in the first place.  In any case, the
ionised pure reflection model does not fit the soft excess as well as the
thermal plasma model (Table~\ref{fit_comparisons}).  Modifications to the
reflection model (such as including contributions from S, Ar and Ca) could
improve the fit to the data.

\subsection{GRB\,010220}

\begin{figure}
 \includegraphics[angle=-90,width=0.93\columnwidth,clip=]{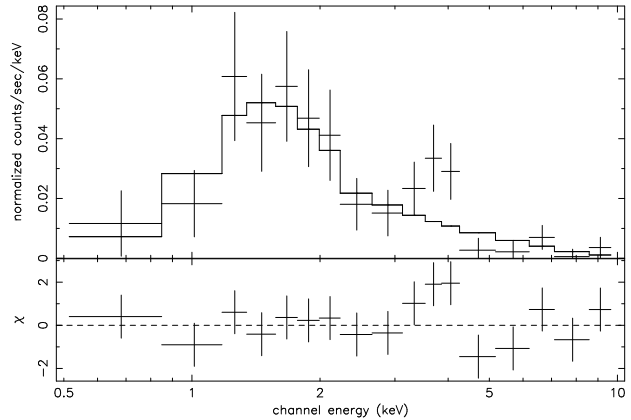}
 \caption{EPIC-pn spectrum of the afterglow of GRB\,010220, 20\,ks exposure,
          fit with a power-law with Galactic absorption.}
 \label{complete_pow_pn_010220_spectrum}
\end{figure}

A power-law model with Galactic absorption is an acceptable fit to the data,
however the single clearest deviation in the spectrum of GRB\,010220
(Fig.~\ref{complete_pow_pn_010220_spectrum}) is the feature near 3.9\,keV;
adding an unresolved Gaussian line (equivalent width $ = 1.8^{+0.8}_{-1.2}$\,keV)
to the power-law fit improves the fit at
$>99$\% significance ($\chi^2 = 7.5$ for 13 degrees of freedom) and is the
only emission feature in the spectrum.  While the relatively poor statistics
do not enable us to detect individual emission lines at lower energies (e.g.
Si or S\,K$\alpha$), the thermal plasma model is significantly preferred to
the absorbed power-law model with a null hypothesis probability of 0.0016
using statistics from the Monte Carlo simulations. Allowing the abundance of
Fe to vary did not significantly improve the thermal plasma fit as the
energy of the Fe K$\alpha$ emission is too low relative to the emission at
low energies and is not broad enough to fit the excess alone.  The best-fit
Fe abundance is consistent with solar abundance; allowing Ni to vary
improves the fit at $>99$\% confidence as it fits most of the 3.9\,keV
emission feature.

The pure ionised reflection model in this case is a better fit than the
absorbed power-law model (Table~\ref{fit_comparisons}) primarily due to
fitting Fe emission (at lower redshift than the thermal plasma model) to the
$\sim3.9$\,keV feature, however the addition of any power-law continuum
disimproves the fit as for GRB\,001025A.

\section{Discussion\label{discussion}}
It is clear that for the X-ray afterglows of GRBs\,001025A and 010220, the
type of collisionally-ionised plasma model fit to the afterglow of GRB\,011211
\citep{2002Natur.416..512R} is a better fit to the data than an absorbed
power-law.
The detection of a thermal spectrum in three GRB afterglows with
\emph{XMM-Newton} indicates that this behaviour is relatively common hours
to days after the burst among the long duration GRBs.  Thermal emission may
therefore make a significant contribution to the total afterglow luminosity
of a GRB.  It is worth noting that these are among the best quality soft
X-ray spectra of GRB afterglows recorded to date and the absence of
detection of such thermal spectra with \emph{BeppoSAX} is not suprising
given \emph{XMM-Newton}'s much greater sensitivity and better spectral
resolution and the relatively high low-energy cut-off of the MECS instrument
($\sim2$\,keV).  A re-examination of the \emph{ASCA}, \emph{BeppoSAX} and
\emph{Chandra} data of the afterglows of GRBs 970508, 970828 (where the
5.04\,keV feature could be caused by blueshifted $^{56}$Ni K$\alpha$ emission
blueshifted by 0.2c), 991216, 990705 and 000214 may be worthwhile in the
light of these results.  

In none of the three cases (GRBs 001025A, 010220 and 011211) is an excess
abundance of Fe required and in both afterglow spectra reported here a large
over-abundance of Ni (or possibly Co) is required, consistent with the idea
that the SN--GRB lag must be of the order of days, not months, though it is
worth noting that this timescale can be extended where the material is
ionised \citep{astro-ph/0205321}.

Using the model proposed by \citet{2002Natur.416..512R} to explain the
observations of GRB\,011211 it is difficult to place precise limits on the
size of the SN shell, the expansion speed and the time delay between the
burst and the initial SN from the spectra of these afterglows as 
the redshifts of the host galaxies are unknown and the redshifts derived
from the X-ray spectra are not as tightly constrained as in GRB\,011211.  It
is possible however, to constrain the minimum radius of the SN ejecta shell
and its mass on the assumption that the duration of the thermal emission is
due primarily to light-crossing time over the illuminated shell.  With no
collimation of the GRB and where the thermal emission lasts at least as long
as the time between the burst and the afterglow observation ($\sim 7$ and 30
hours), the minimum shell radii are $8\times10^{14}$\,cm and
$3\times10^{15}$\,cm corresponding to SN-GRB delays of at least 3 and 12
days for GRB\,010220 and GRB\,001025A respectively and illuminated shell
masses of 0.9 and 1.1\,M$_{\sun}$ using the best fit emission measure,
temperature and redshifts from the MEKAL model.
\citet{2002Natur.416..512R} calculate a delay of a few days and at most two
weeks between the SN and the GRB, consistent with these results, though the
SN-GRB time delay distribution has so far not been defined. The time
delay for GRB\,001025A is suggestive since the half-life of the most
abundant product of Si burning, $^{56}$Ni, is 6.1 days decaying primarily to
$^{56}$Co with a half-life of 77.3 days to $^{56}$Fe 
\citep[although see][on half-lives of ionised species]{astro-ph/0205321}. 
In this scenario, much of the emission observed near 1\,keV in the spectrum
of GRB\,001025A could then be due to L-shell emission from $^{56}$Co.  The
spectrum however is not sufficiently good to investigate this possibility.

Recent analysis of the \emph{XMM-Newton} spectrum of GRB\,020322, with a
signal to noise ratio comparable to the X-ray spectrum of GRB\,011211, shows
no evidence of emission features (Watson et~al. in prep.) 15\,hours after the
burst; evidence that if thermal emission occurs in every afterglow, it may
be relatively short-lived.

The prospect that thermal emission may contribute significantly to the X-ray
luminosity of afterglows could resolve the problem of bursts with apparently
high X-ray column densities but low optical extinctions, 10--100 times
smaller than expected \citep{2001ApJ...549L.209G}.  The curvature of
low-quality thermal X-ray spectra can be approximated by an absorbed
power-law, giving much larger absorbing columns than are required by the
thermal spectra.

Another implication of the data reported here is that GRBs without
significant optical emission, ``dark GRBs''
\citep{astro-ph/0107545,2001ApJ...562..654D}, may not always need
to have significant extinction
\citep{2000ApJ...537L..17T,astro-ph/0107545,astro-ph/0201282}, be
intrinsically different \citep{2002MNRAS.330..583L} or be very distant
\citep{2000ApJ...536....1L}.  A soft X-ray spectrum dominated by thermal
emission will lead to an overestimation of the expected optical flux where
this is predicted assuming a synchrotron-dominated X-ray flux below 10\,keV
\citep{2001ApJ...562..654D}.

\section{Conclusions\label{conclusions}}
\emph{XMM-Newton} detected the afterglows of GRB\,001025A and GRB\,010220
the former showing a decaying lightcurve.  Positions accurate to\ 
$\sim1$\arcsec\ were determined; this makes it feasible for optical spectroscopy
to be used to confirm the X-ray redshift reported in this paper for the host
galaxies of these GRBs, in particular for the host of GRB\,001025A, where
the extinction is relatively low.  In both cases, the X-ray spectra are
significantly better fit by the thermal plasma model proposed by
\citet{2002Natur.416..512R} to explain the line features in GRB\,011211 than
by an absorbed power-law and in the case of GRB\,001025A the thermal plasma
model is a significantly better fit than an ionised reflection model.  The
parameters determined from these thermal fits are consistent with their
suggested scenario.  It seems likely that a large contribution from
highly-ionised light metals is a common feature in the X-ray spectra of GRB
afterglows hours to days after the burst and the detection of thermal
emission in three of the four \emph{XMM-Newton}--detected afterglows implies
that this may be a significant component in the total afterglow luminosity
of all long GRBs.

\bibliography{mnemonic,grbs}

\end{document}